\documentclass[aip,twocolumn,amsmath,amssymb,reprint]{revtex4-1} \synctex=1 

\usepackage{graphicx}
\usepackage{dcolumn,array}
\usepackage[colorlinks=true,linkcolor=blue]{hyperref}
\usepackage{hyperref}
\usepackage{bm}

\begin{document}

\preprint{APS/123-QED}

\title{Flow and clogging behavior of a mixture of
  particles in a silo}

\author{Sukhada C. Bhure} 
\affiliation{CSIR-National Chemical Laboratory, Pune 411008 India} 
\affiliation{Academy of Scientific and Innovative Research (AcSIR), Ghaziabad 201002 India}

\author{Pankaj Doshi} \email{pankaj.doshi@pfizer.com}
\affiliation{Pfizer Research and Development, Pfizer Products India Private Limited, Mumbai 400051, India}

\author{Ashish V. Orpe} \email{av.orpe@ncl.res.in} 
\affiliation{CSIR-National Chemical Laboratory, Pune 411008 India} 
\affiliation{Academy of Scientific and Innovative Research (AcSIR), Ghaziabad 201002 India}

\date{\today}

\begin{abstract}
  We investigated the clogging behavior observed during the flow of
  aspherical particles from a silo in the presence of spherical
  particles of different sizes and proportions using
  flow visualization experiments and discrete element method (DEM)
  simulations. The size of the avalanche, essentially the tendency of
  clogging, exhibits non-monotonic dependence on the spherical
  particle volume fraction. For small enough content of spherical
  particles, the clogging tendency intensifies, whereas it reduces
  rapidly for high enough spherical particle fractions, with a minimum
  in between. The non-monotonic behavior is observed to persist over for
  different spherical particle sizes. The overall behavior is
  shown to arise due to competing effects between the localized total
  particle fraction influencing avalanche strength and mean size of
  the particles exiting the silo, influencing the probability of arch formation.
\end{abstract}

\maketitle

\section{Introduction}

The phenomena of clogging during the outflow of dry granular material
from a silo, while interesting in itself, has also resulted in
throwing up another interesting phenomena of unclogging of the clogged
silo. While the former has been studied over several
years~\cite{to01,tang11,tewari13,thomas13,zuriguel14b,thomas16}, the
latter has garnered attention in recent times. It has been shown that
the silo unclogging may be forced through air jet impinging or silo
vibration~\cite{janda09b}, by having multiple exit
orifices~\cite{kunte14,orpe19}, or by placing inserts at suitable
locations inside the silo~\cite{zuriguel11}. These forcings can either
break the stable arch or reduce the probability of arch formation in
the first place, which is primarily responsible for
clogging. Interestingly, the clogging-unclogging phenomena have also
been extended to natural systems like movement of pedestrians or
animals through a narrow exit~\cite{zuriguel14a,zuriguel17} or
artificial systems like the flow of colloidal particles through an
orifice~\cite{hidalgo18}.

Recently, it has been shown that the flow in a (2-dimensional) silo
can be enhanced and clogging tendency be reduced due to presence of
other (secondary) particles, which are smaller than the bulk
particles, but are present at volume fractions as high as
$0.2 - 0.4$~\cite{nicolas18b,madrid21}. The silo was operated in the
vibration mode, presumably to well mix two different sized
particles. The probability of clog formation was found to be reduced
due to the presence of secondary particles which led to the formation of
an arch easily breakable due to vibration~\cite{nicolas18b}. The
presence of these particles also showed an increase in the overall
flow rate with an optimal dependence on the particle
size~\cite{madrid21}. Our interest lies in understanding the effect
of such secondary, smaller spherical particles on the
clogging/flow behavior in the silo, but in small amounts,
akin to the presence of trace impurities in the flow. In practice, the
impurities are bound to be present in the powder material, the effect of
which on the flow or clogging is invaluable. Furthermore, we intend to
study the flow and clogging behavior of aspherical particles,
encountered mostly in practical situations. For simplicity, we
consider the secondary particles to be spherical in shape.  Apart
from practical considerations, the study of the flow of non-spherical
particles is fundamentally interesting in its own way. They are known
to exhibit clogging characteristics differing from those observed for
spherical particles. For instance, the breakdown of the exponential
trail in avalanche distributions~\cite{zuriguel05}, higher probability
to form clogs due to possibility of multiple contact points and
approach to spherical particle behavior with increased vertices in a
polygon~\cite{goldberg15,goldberg17} are some of the peculiar observations
related to non-spherical particle shape.

In this work, we focus on understanding the clogging and flow behavior
of cylinder-shaped particles in the presence of small amounts of
spherical particles in a 3-dimensional silo. We carry out the required
study using flow visualization experiments and discrete element method
(DEM) simulations. In Sec.II, we describe the experimental
system and simulation details followed by results focusing on
the avalanche size behavior and relevant characteristics of the
system. Toward the end, we provide quantitative reasoning to explain the
observed phenomena.

\section{\label{method}Methodology}
\subsection{\label{expt}Experimental details}

\begin{figure} \includegraphics[scale=0.5]{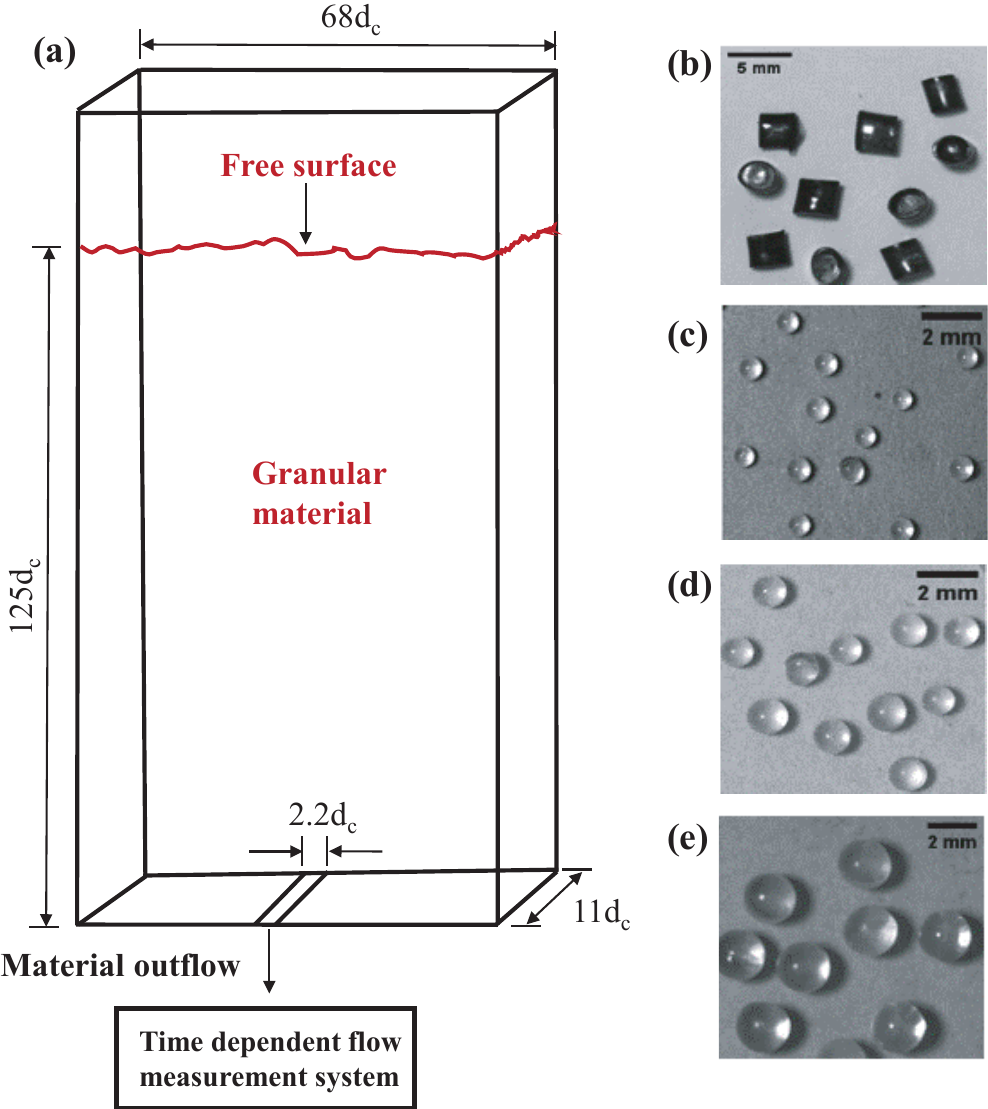}
  \caption{(a) Schematic of experimental silo system (b) Cylinder-shaped bulk particle with equivalent sphere volume diameter
    $d_{c} = 3.2$mm.  Spherical particles with diameters ($d_{s}$) of
    $0.7$mm, $1.0$mm and $2.0$mm are shown, respectively, in (c),
    (d) and (e).}
  \label{schem-expt}
\end{figure}

Experiments are performed in a silo with dimensions as shown in
Fig.~\ref{schem-expt}.  The side and bottom walls of the silo are made
out of acrylic plates glued to each other, while the top is kept open
to pour the granular material. The material outflows from an exit slit
of fixed width. Two types of particles are used in the
experiment. The bulk particles comprise of cylinders made from
poly-methyl methacrylate (PMMA), of density $1.2$g/cc, with an
elliptical cross-section. The length of each cylinder is $3.1$mm,
while the major and minor diameters are of length $3.2$mm and $2.2$mm, respectively. The diameter ($d_{c}$) of an equivalent sphere
volume is, then, calculated as $3.2$mm. Glass beads, of density $2.5$g/cc
and three different diameters ($d_{s}$), viz., $0.7$, $1$ and
$2$mm are used as secondary spherical particles resulting in
particle size ratios ($r = d_{c}/d_{s}$) of $4.6$, $3.2$ and $1.6$
respectively. Images of these particles are shown in
Figs.~\ref{schem-expt}(b)-\ref{schem-expt}(e).

For each experiment, the silo was filled with a mixture of bulk and
one of the spherical particles up to a height of $125 d_{c}$, while keeping the
exit slit closed. The mixture was prepared by manual mixing of bulk
cylindrical particles with predefined spherical particle volume
fraction ($\phi$) varied in the range $0$-$0.1$. The mixture was,
then, poured in the silo by employing distributed filling method. The
outflow from the silo was initiated by opening the exit orifice, and
the flowing material was collected directly on a weighing scale. Given
the size of the orifice with respect to the bulk particle size, the
orifice clogged after flowing for some time. The total mass,
comprising both the particles, collected on the weighing scale
till the occurrence of clogging was termed as the avalanche size
($S$). After a wait time of $5$s, the flow was reinitiated by
piercing the exit arch with a pointed object to trigger another
avalanche. The procedure was repeated till the fill height reduced to
$50d_{c}$, following which the remainder of the material was emptied out,
and the silo refilled back to a height of $125d_{c}$ to restart the
experiments. For every spherical particle concentration and size
employed, the silo was filled about five times resulting in at least
$500$ independent avalanche events. The above procedure was repeated
for different volume fractions and sizes of spherical beads.

\subsection{\label{sim}DEM simulations}

\begin{figure} \includegraphics[scale=0.5]{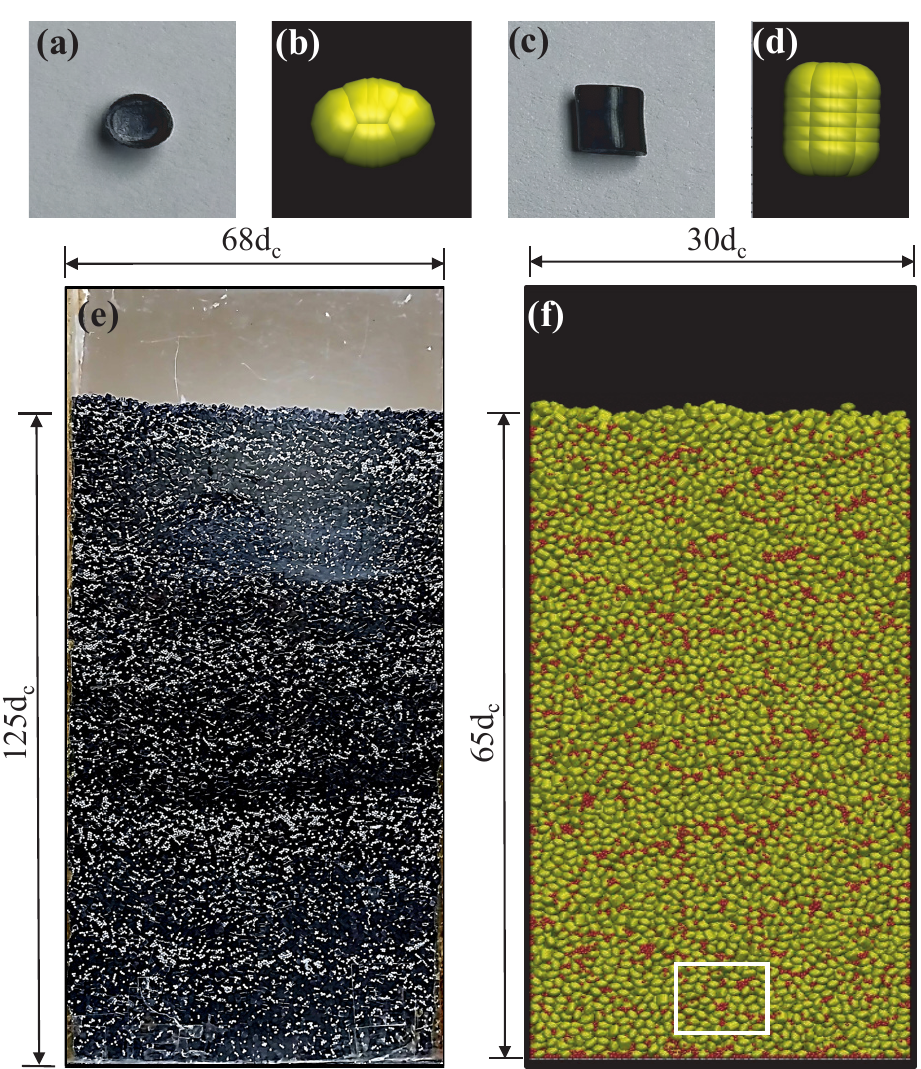}
  \caption{Simulated particle and silo system. (a) Top view of
    the experimental particle. (b) Top view of the simulated particle. (c)
    Front view of the experimental particle. (d) Front view of the simulated
    particle. (e) and (f) represents the front view of the silo filled
    with a binary mixture of cylindrical and spherical particles,
    respectively in experiments and simulations, with $\phi = 0.05$
    and $r = 3.2$.  The white box near the exit represents the region
    of interest for the requisite analysis as described later in the text.}
  \label{schem-sim}
\end{figure}

The discrete element method (DEM) simulations were carried out to
investigate the clogging of aspherical particles in the presence of
secondary spherical particles. The simulations were carried out using
an open source program, LIGGGHTS (LAMMPS Improved for General, Granular
and Granular Heat Transfer Simulations), for different sphere volume
fractions. The non-spherical particle (cylinder with elliptical
cross section) was created using the in-built ``multisphere'' routine
of the LIGGGHTS software. The approach involves clumping together
predefined number of spheres of specific size. The resulting
non-sphericity of the particle is dependent on the number of clumped
spheres, their location with respect to each other and the degree of
overlap. Over here, we considered $50$ spheres to generate a cylinder
with elliptical cross section, while maintaining the relative
magnitudes of the length, minor diameter, and major diameter the same
as for the experimental particle. As seen from Figs.~\ref{schem-sim}(b)
and \ref{schem-sim}(d), the shape of the particle generated in simulations seems to be
reasonably close to the experimental particle [see
Fig.~\ref{schem-expt}(b)], barring the sharp corners. A further
increase in the number of spheres allowed for replicating the corners
in a better manner, but required much longer simulation time. However,
this did not alter the flow behavior significantly and, hence, was not
considered.  The secondary spherical particle was simply modeled as a
sphere of prescribed size.

The silo of smaller size, viz. height $65d_{c}$, width $30d_{c}$, and
depth $10d_{c}$, was used in simulation. Reduced dimensions of the silo
were used to reduce the simulation time while ensuring the absence of
wall effects. Moreover, the number and the size ratio of aspherical
particles to spherical particles was kept the same as in experiments
for a given sphere volume fraction. Figures \ref{schem-sim}(e) and \ref{schem-sim}(f) 
show the front view of the silo, respectively for experiments and
simulations, comprising a mixture of particles ($\phi = 0.05$) for a
particle size ratio $r = 3.2$. 

The simulation employs Hertzian contact model for calculation of force
between two contacting particles. The Hertzian contact model was used
in this work due to its ability to better capture the realistic
behavior of granular materials, given the dependence of the interaction
force on the overlapping area instead of overlapping distance
considered in Hookean contact models~\cite{thornton11,yeom19}. The
Hertzian model, thus, results in a more realistic force evolution and
has been used 
previously quite
frequently~\cite{thornton11,markaus11,liu14,yeom19,tangri19}.  The
contact force comprises of normal ($F_{n}$) and tangential ($F_{t}$)
components, each of which includes two terms given as
\begin{equation}
  \bm{F_{n}} = \left (k_{n} \delta \bm{n} - \frac{\gamma_{n}
      \bm{v}_{n}}{2} \right),
\end{equation}
\begin{equation}
  \bm{F_{t}} = -\left (k_{t} \Delta \bm{s}_{t} + \frac{\gamma_{t}
      \bm{v}_{t}}{2} \right),
\end{equation}
where $\bm{n}$ is the unit vector along the line connecting centers
of two particles, $\bm{v}_{t}$ and $\bm{v}_{n}$ are, respectively, the
tangential and normal components of particle velocities.  Both, the
normal elastic constant ($k_{n}$) and tangential elastic constant
($k_{t}$) are chosen to be of the order of $10^{7}$ $mg/d_{\alpha}$.
The values of the normal damping term ($\gamma_{n}$) and tangential
damping term ($\gamma_{t}$) are chosen to be of the order of
$10^{2} \sqrt{g/d_{\alpha}}$. Here, $d_{\alpha}$ represents either
$d_{c}$ for cylindrical particles or $d_{s}$ for spherical particles,
and $g$ represents gravity acting in downward direction.  The value
$\Delta \bm{s}_{t}$ is the tangential displacement between two
particles to satisfy the Coulomb yield criterion given by
$\bm{F_{t}} = \mu_{s} \bm{F_{n}}$, where $\mu_{s}$ is the coefficient
of static friction coefficient.  The density ratio between cylindrical
bulk particles and spherical particles was maintained the same as in
experiments. The integration time step used in the simulation is
$10^{-4}$.  The silo was filled with mixture up to a height of $65d_{c}$,
and the flow in simulations was initiated by opening the exit
slit. The total mass of particles flowing out of the silo before the
orifice clogged was termed as the avalanche size ($S$). The flow was
reinitiated by removing a few particles in the arch. This procedure was
repeated till the fill height reduced to $40d_{c}$. This resulted in
about $70$ independent avalanche events. However, unlike that in
experiments, the silo was not refilled to record more number of
avalanche events as that became computationally prohibitive. However,
as shown later, about $70$ recorded avalanche events per case seems
reasonably large enough to capture the observed experimental
phenomena.

\subsection{\label{param}Parameter calibration}

\begin{table}
  \caption{\label{frictionval} Values of coefficient of static and
    rolling frictions for different contact pairs used in simulations}
  \begin{ruledtabular}
    \begin{tabular}{lrr}
      Contact pair & $\mu_{s}$ & $\mu_{r}$ \\ \hline
      Cylinder-cylinder & 0.9 & 0.2 \\
      Cylinder-sphere & 0.7 & 0.15 \\
      Sphere-sphere & 0.4 & 0.0 \\
    \end{tabular}
  \end{ruledtabular}
\end{table}

\begin{figure}[t] \includegraphics[scale=0.3]{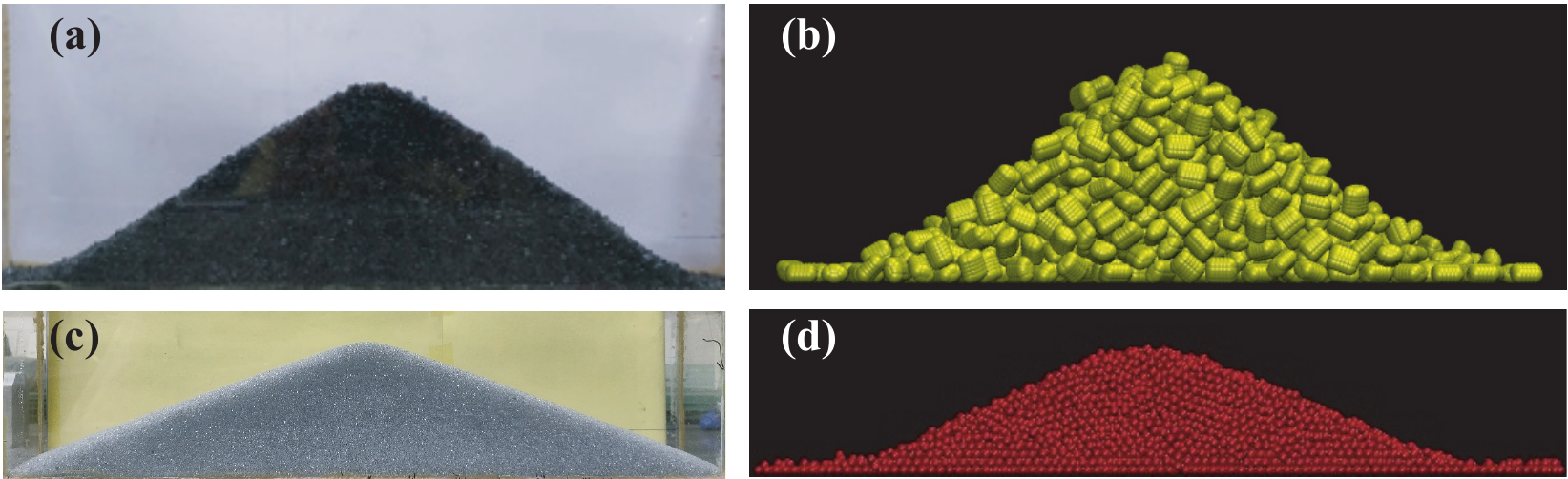}
  \caption{Snapshot of the granular pile created for cylindrical
    particles in (a) experiments and (b) simulations and for spherical
    particles in (c) experiments and (d) simulations for
    particle size ratio ($r$) of 3.2.}
  \label{calibr}
\end{figure}

The calibration of the simulation parameters to match the experiments
can be an uphill task given the range of parameters used and two types
of particles and materials employed. For simplicity, except friction
coefficients, all the remaining contact model parameters are of the
same order of magnitude, typically as used
previously~\cite{rycroft09,arevalo16}, but for glass
beads~\cite{silbert01,rycroft09,orpe19} which has a modulus about an
order of magnitude greater than PMMA. Note that these values of the
contact model parameters are substantially lower than those relevant
to real glass and are chosen so as to reduce the overall computational
effort~\cite{landry03,orpe19}. In addition to the coefficient of
static friction ($\mu_{s}$), we have also employed coefficient of
rolling friction ($\mu_{r}$) between the particles. The latter
represents the ease with which the particles roll past one another and
will be determined by the asphericity of the particles in
contact. Typically, for the sphere-sphere contact, the value can be
expected to be quite low, while it can be high for the cylinder-cylinder
contact.  The values of both the friction coefficients were adjusted
so as to match the value of static angle of repose measured from
simulations to that measured from experiments.

The material, either cylinders or spheres, were slowly poured in a
rectangular cell to form a static heap. The length, height, and depth
of the experimental cell was $150d_{c}$, $65d_{c}$, and $18d_{c}$,
respectively, while that of simulation cell was $30d_{c}$, $15d_{c}$,
and $10d_{c}$, respectively.  The cell dimensions were sufficiently
large to prevent any kind of end effects. In experiments, the static
image of the heap was captured using a digital camera positioned
sideways and orthogonal to the sidewall of the rectangular cell [see
Figs.~\ref{calibr}(a) and \ref{calibr}(c)]. In simulations, the angle was measured from
the final static position of the particles exported as an image [see
Figs.~\ref{calibr}(b) and \ref{calibr}(d)]. In each image, a central, nearly flat free
surface region of the heap (about $15d$) was analyzed to obtain the
angle of repose. Every experimental measurement was repeated about six
times to ensure consistency. The values of friction coefficients
reported in Table~\ref{frictionval}, averaged over all spherical
particle sizes, correspond to the scenario wherein the angle measured
for an experimentally prepared pile varies within half a degree from that
measured in a pile from simulations. This close agreement suggests
that the finalized simulation parameters seem reasonable enough to
qualitatively reproduce experimental observations and were used in
carrying out simulations of particles flowing out of silo.

\section{\label{results}Results \& Discussion}

In the following, we first provide a qualitative understanding from
the observations. This is then followed by a quantitative discussion
of the clogging behavior in terms of avalanche size variations as
observed in experiments and compared with those observed in
simulations. Toward the end, we discuss about certain specific
measurements from the simulation data, which have been used to explain
the experimental observations.

\begin{figure*} \includegraphics[scale=0.45]{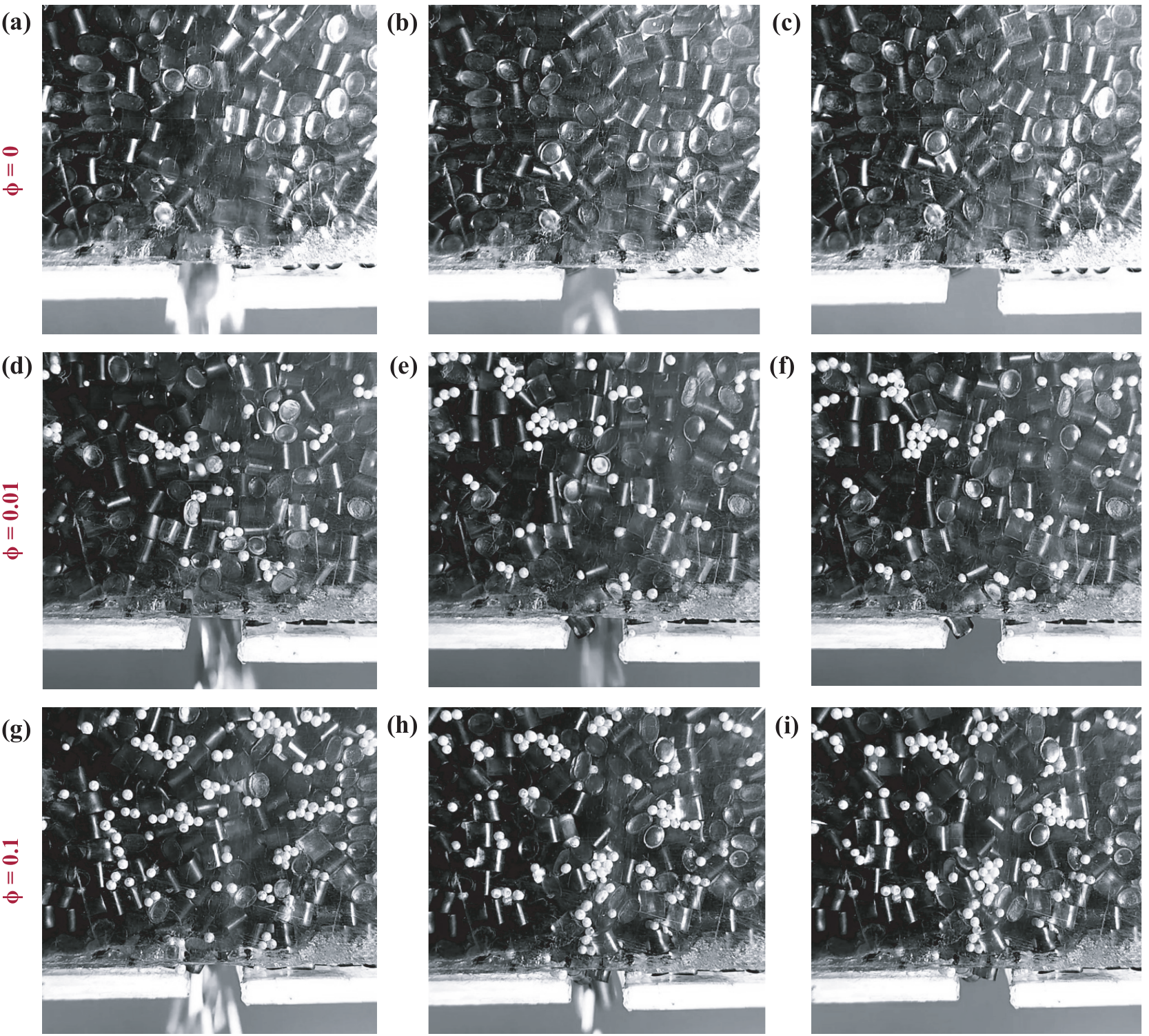}
  \caption{Images from experiments
    showing particle configurations during flow and arch formation in
    the vicinity of the silo exit for a few different fractions of
    spherical particles ($r = 3.2$).  Spheres are seen as white
    circles in the images. All the images are acquired at the front
    wall.}
  \label{archvisexp}
\end{figure*}

\begin{figure*} \includegraphics[scale=0.47]{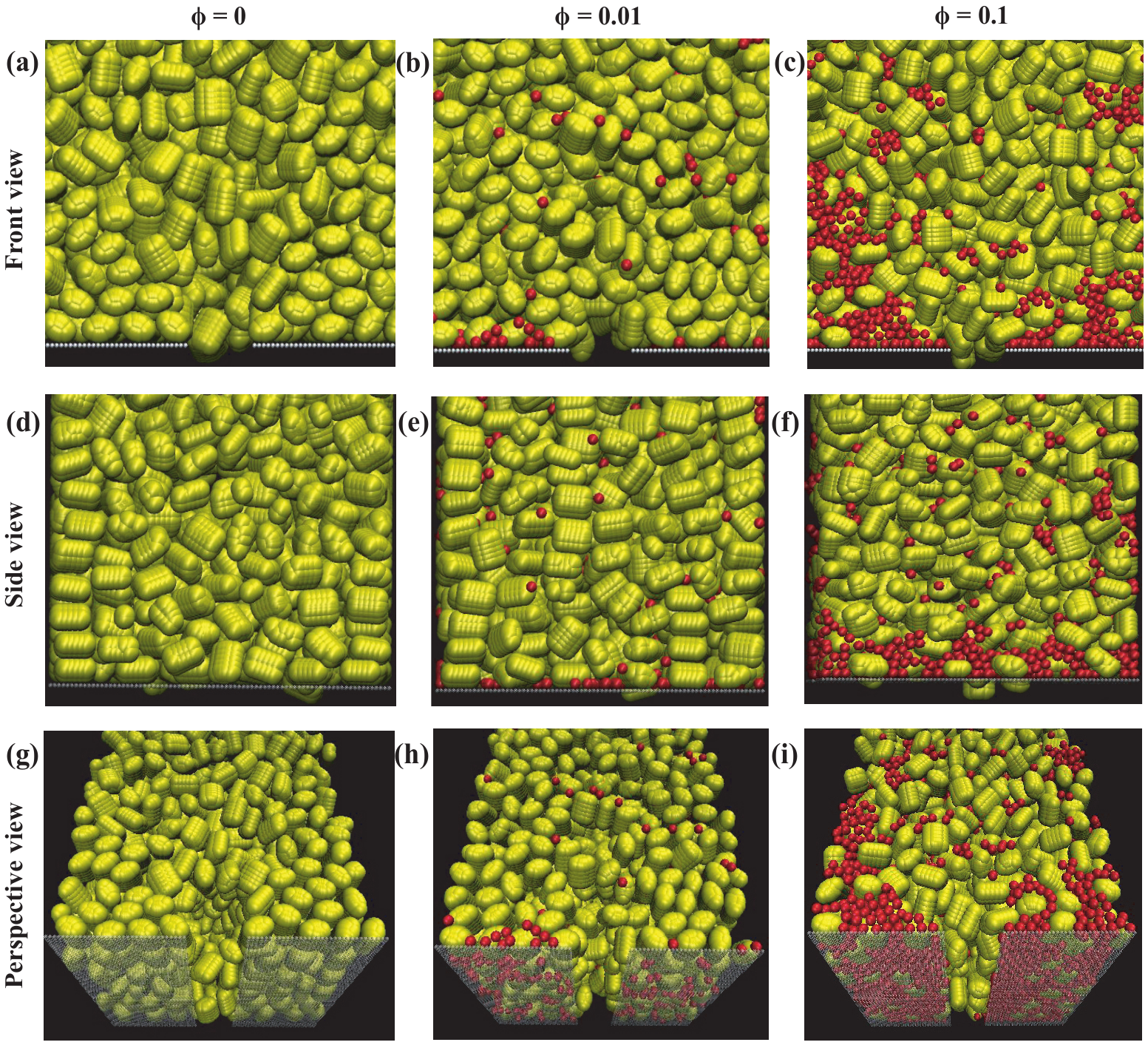}
  \caption{Images from simulations
    showing particle arrangements in the vicinity of the silo exit for
    a few different fractions of spherical particles ($r = 3.2$) after
    the flow has clogged. Images in the first panel (a-c) are acquired
    near the front wall. Images in second panel (d-f) are acquired
    near one of the side walls. Images in the third panel (g-i) are
    acquired from the bottom at a perspective angle to the silo. The
    shaded region in the third panel represents the bottom plate with
    a slit in between.}
  \label{archvissim}
\end{figure*}

Figure~\ref{archvisexp} shows images
acquired in experiments during an avalanche at different times and for
three different volume fractions ($\phi$) of spherical particles of
diameter $d_{s} = 1$ mm, i.e., $r = 3.2$. The images show particle
configurations at different stages of an avalanche in the vicinity of
the exit orifice near the front wall of the silo. The video
representation of these images for different sphere volume fractions
and sizes is available as a supplementary material. In each case, as
seen from the respective videos, the overall motion appears
intermittent, eventually leading to an arch formation and clogging. The
number of small spheres increase with increasing sphere concentration,
though the distribution is not uniform. The spheres, appearing as
distinct and small clusters, seem to fill in the voids created by the
packing of cylinders. The number of clusters increase with increasing
packing fraction, though the size of the cluster remains nearly the
same. This is suggestive of the upper limit for the void space
available for filling due to smaller spherical particles. The
formation of the arch and its shape does not seem to depend on the
spherical particle fraction, underlining the random nature of the
event in all cases. Near identical, qualitative behavior is also
observed for other spherical particle fractions and sizes.

The simulation counterpart for the earlier discussion on experimental
observations is shown in fig.~\ref{archvissim}.  The images are acquired from three different view positions
to understand the 3-dimensional nature of the flow and clogging. Unlike in
fig.~\ref{archvisexp}, the images in
fig.~\ref{archvissim} are only shown for
the final clogged state. The flow of the avalanches leading to
clogging, however, can be seen from the videos available as the
supplementary material for different particle fractions and sizes. The
main qualitative features seen in experiments, i.e., increased number
of smaller spheres with increased fractions, formation of clusters are
observed in simulations too.  There is, however, a hint of the
occurrence of segregation near the bottom, not seen clearly in
experiments. The corresponding view from the sidewall (second panel
in the figure) shows similar configuration of spherical particles and
cylinders showing that the behavior is not a localized, but rather a bulk
phenomenon. The perspective view in third panel in
Fig.~\ref{archvissim} shows the complicated nature of the arch across
the depth of the silo. The occurrence of the flow even when the
particles near the front wall are stationary is a direct consequence
of this 3-dimensional nature of the arch formation.

While the images in figs.~\ref{archvisexp} and \ref{archvissim} and
corresponding videos show the behavior for one spherical particle
size, a similar qualitative behavior is also observed for other
spherical particle sizes. From all these observations, it can be
anticipated that the presence of spherical particles in the voids,
individually or in clusters, may influence the flowability of the
system and arch forming tendency. However, these can be quantitatively
ascertained by measurement of the mean and distribution of the
avalanche sizes and their dependence on spherical particle size and
fractions as discussed next.

\begin{figure} \includegraphics[scale=0.6]{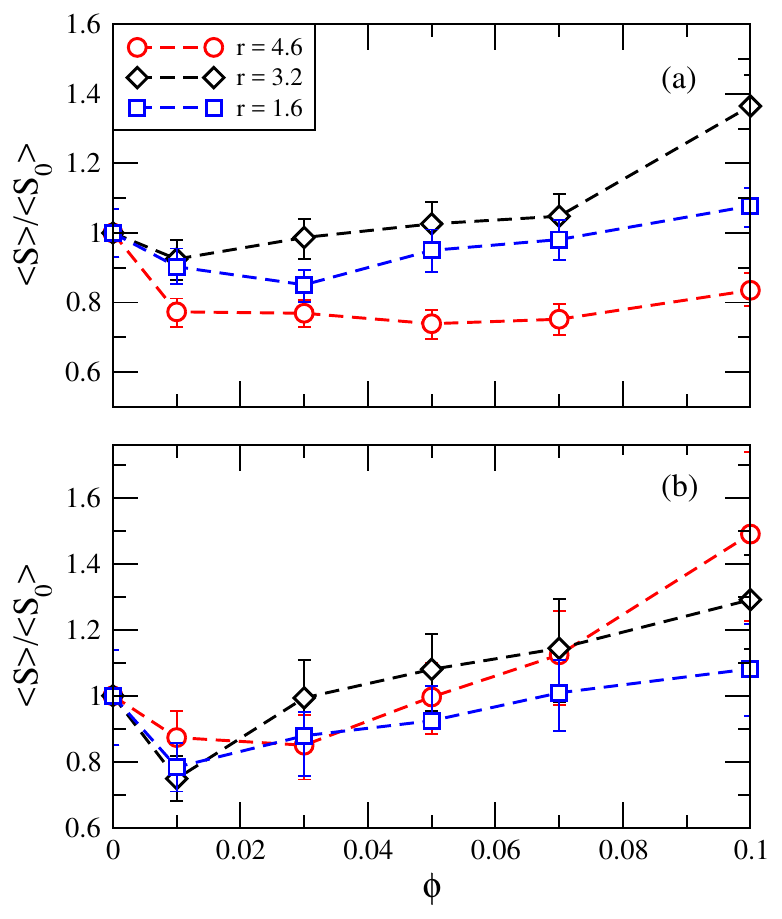}
  \caption{Variation of normalized average avalanche size
    ($\langle S \rangle/\langle S_{0} \rangle$) with spherical particle
    volume fraction ($\phi$). Data from (a) experiments and (b)
    simulations for three different particle size ratios ($r$).
    $\langle S_{0} \rangle$ represents average avalanche size in the absence
    of spherical particles. Error bars represent confidence interval of $95$\%.}
  \label{clog-behv-expt}
\end{figure}

The avalanche size represents the amount of material flowing out of
silo till the orifice is clogged. The avalanche size ($S$) in
experiments is the total mass of cylindrical and spherical particles
collected during the flow before the occurrence of clog. The average value
($\langle S \rangle$) is obtained over $500$ independent flow (or
clogging) events. The variation of normalized average avalanche size
($\langle S \rangle/\langle S_{0} \rangle$) with the spherical
particle concentration ($\phi$) is shown in fig.~\ref{clog-behv-expt}(a). Here,
$S_{0}$ represents the avalanche size for the base case, i.e., in the
absence of spherical particles ($\phi = 0.0$). Several interesting
features are evident from this figure, which we dwell upon next.

The magnitude of the avalanche size is governed by the ability of the
flowing particles to form a stable arch. The increased avalanche size
represents longer flow duration before clogging takes place, indicating
a lesser tendency to form a stable arch and vice versa for decreased
avalanche sizes. In the limit of infinite avalanche size (when flow
never stops), the tendency of arch formation will be negligible, and, in
the limit of no flow or immediate clogging, the tendency will be quite
high. It may be intuitively expected that the presence of small
spherical particles may lubricate the flow of cylindrical
particles and/or may reduce direct contacts between cylinders thereby
reducing the tendency of arch formation, resulting in increased avalanche
size. However, as seen from Fig.~\ref{clog-behv-expt}(a), the observed
behavior is exactly the opposite. The presence of small spherical
particles actually reduces the avalanche size compared to the base
case, essentially aiding the clogging behavior. The decrease in the
avalanche size is, however, observed over a limited range of sphere
volume fraction, leading to a minimum in the avalanche size at an
intermediate sphere volume fraction, which is dependent on the size of 
spherical particle. Any further addition of spherical
particles increases the avalanche size, i.e., inhibits clogging tendency.

For higher values of $\phi$, the avalanche size seems to increase
rapidly, similar to that observed
previously~\cite{nicolas18b,madrid21} This can be expected as, with
increasing volume fraction of spherical particles in the system, the
proportion of these particles within the total number of particles
exiting the silo will also increase. Given much larger orifice size
relative to the size of spherical particles, they cannot be expected
to form an arch. Moreover, their presence in large numbers will also
reduce the mean size of particles exiting the silo, thereby reducing
the tendency of forming an arch, the result being a larger avalanche
size. Indeed, the avalanche size will diverge at high enough spherical
particle concentration approaching $\phi = 1.0$, wherein there will be
primarily spherical particles in the outflow thereby precluding arch
formation and hence no clog formation.  While the overall
(non-monotonic) behavior remains same for different sizes of spherical
particles, certain deviations exists, which are not quite systematic
and hence difficult to understand at this time. For instance, the
minimum avalanche size spreads over a range of spherical particle
fractions for smallest size spherical particle ($r = 4.6$) [(see
Fig.~\ref{clog-behv-expt}(a)], while the spread is limited to a narrow
range of fractions for the other two size ratios.

\begin{figure} \includegraphics[scale=0.6]{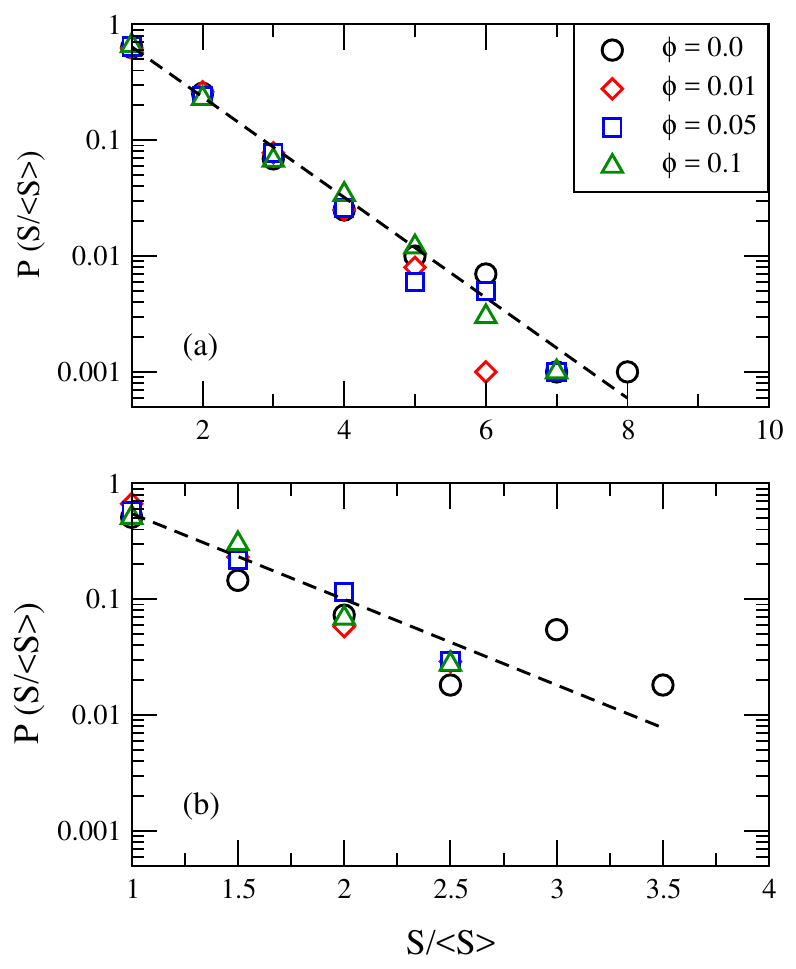}
  \caption{Probability distribution of normalized avalanche size
    ($S/<S>$) for size ratio ($r = 3.2$) and different spherical
    particle volume fraction in (a) experiments and (b)
    simulations. The dashed line represents an exponential fit.}
  \label{aval-dist}
\end{figure}

The equivalent data as acquired from DEM simulations is shown in
Fig.~\ref{clog-behv-expt}(b) for the same three particle size ratios as
in experiments. While the data for $r = 3.2$ and $r = 1.6$ is nearly
the same as that obtained in experiments, qualitative differences are
seen for the case of smallest sphere of $r = 4.6$ compared with
experiments. The more sustained minimum observed in experiments is not
seen in simulations (see Fig. S1 provided in the supplementary
material for better comparative representations). This disparity in
the observations for different spherical particle sizes, in
experiments as well as simulations, can be possibly attributed to the
lack of exact replication of (i) the shape of cylindrical particle in
simulations, particularly the sharp edges and (ii) actual
inter-particle interaction. This inadequacy seems to have a different
effect with respect to spherical particle concentration and size, the
origins of which are not clear at the moment. However, more importantly,
the non-monotonic variation of normalized avalanche size with
spherical particle volume fraction is exhibited for all the three
sphere sizes, suggestive that the physics behind the experimental
observations is captured quite well in simulations. In that case, the
more detailed and the three dimensional simulation data can, then, be
used for explanation of the observations as discussed later.

The distributions of avalanches exhibit exponential behavior for all
the cases as shown in Fig.~\ref{aval-dist}. The exponential decay
represents random nature of discrete avalanche events as has been
shown previously in previous
studies~\cite{zuriguel05,janda09b,kunte14,zuriguel14b,orpe19}. This
suggests that the presence of spheres does not influence the inherent
random nature of the clogging phenomena. However, the length scale of
the exponential decay is different for experiments and
simulations. Near-similar qualitative behavior is obtained for
remaining sphere volume fractions as well as for different spherical 
particle sizes.

As discussed earlier, the increase or decrease in the avalanche size
will be governed by the probability of clogging occurrence,
which in turn will depend on probability of arch formation subject to
the local flow conditions prevailing near the silo exit. In a recent
work~\cite{arevalo16}, it was shown that the clogging occurrences
reduced monotonically with increased translational kinetic energy in
the system,
which was increased by increasing the driving (gravity) force and
increasing the orifice width. The authors assumed that for small
enough translational kinetic energy, i.e., slow enough flow, the formed arch is
able to resists its breakage till the flow eventually stops. The exact
reverse happens for fast flows, wherein the arch is unable to prevent
its breakage, thereby reducing the chances of clogging.

We borrow the same argument over here to explain our
observations. However, the possible drivers for altering the kinetic
energy over here are the spherical particle concentration and size
ratio, while maintaining the gravitational force and orifice width
constant. Under these circumstances, the only possible way for
increase or decrease in the kinetic energy is the variation of the
packing fraction in the system. We have calculated the packing
fraction during the flow in a box of length $10d_{c}$, width $10d_{c}$,
and depth $10d_{c}$, located about $3.5d_{c}$ above the exit orifice
[shown in Fig.~\ref{schem-sim}(f)]. The location of the box represents
the region closest to the orifice, which can be expected to exert
maximum influence, while also away from the actual arch formation
location, which typically ranges in the region up to $3d_{c}$ above the
orifice. The volume fraction in the box was obtained as an average
over all the particles across all avalanches and also over all times
within each avalanche. It is to be noted that over the entire
avalanche duration, the volume fraction measured within the region
varied up to $0.5$\% of the initial value when the avalanche was
initiated. Thus, the average value reflects the packing state during
the flow for a specified spherical particle concentration and size
ratio, which, we believe, should influence the average kinetic energy
of the system. We consider both, the translational component of
kinetic energy ($k_{te}$), which accounts for the flow speed of all
particles, and the rotational component of the kinetic energy
($k_{re}$), which predominantly represents the ability of the
cylindrical particles, to orient themselves appropriately. The latter
quantity can be expected to correlate with the average angle of
orientation ($\theta$) of the cylindrical particles, calculated with
respect to the vertical (or flow direction).

\begin{figure} \includegraphics[scale=0.55]{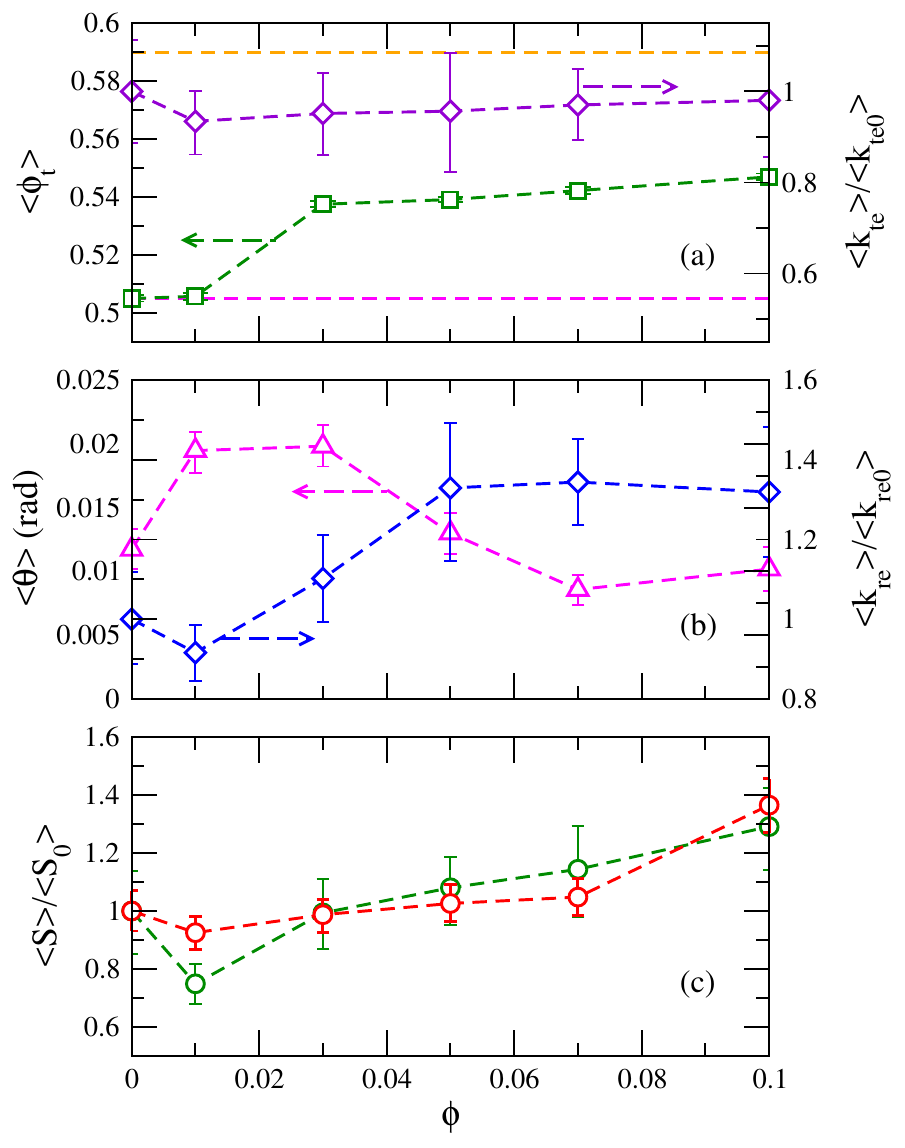}
  \caption{(a) Variation of total (spheres and cylinders) average
    volume fraction ($\langle \phi_{t} \rangle$, green squares) and
    normalized average translational kinetic energy
    ($\langle k_{te} \rangle / \langle k_{te0} \rangle$, violet
    diamonds) of flowing particles with spherical particle volume
    fraction ($\phi$). The dashed magenta colored line and orange
    colored line, respectively, represent the volume fraction during
    flow of only cylinders and only spherical particles. (b) Variation
    of average angle of orientation of the cylinders with vertical
    ($\langle \theta \rangle$, magenta triangles) and normalized
    average rotational kinetic energy
    ($\langle k_{re} \rangle / \langle k_{re0} \rangle$, blue
    diamonds) of flowing particles with spherical particle volume
    fraction ($\phi$) (c) Variation of normalized average avalanche
    size ($\langle S \rangle / \langle S_{0} \rangle$) with spherical
    particle volume fraction ($\phi$) measured experimentally (red
    circles) and in simulations (green circles). The data is obtained
    for a size ratio of $r = 3.2$. Error bars represent confidence
    interval of $95$\%}
  \label{expt-sim-compr}
\end{figure}

Figure~\ref{expt-sim-compr}(a) shows the variation of the average total
volume fraction ($\langle \phi_{t} \rangle$) and normalized average
translational kinetic energy
($\langle k_{te} \rangle / \langle k_{te0} \rangle$), normalized
average rotational kinetic energy
($\langle k_{te} \rangle / \langle k_{te0} \rangle$), and average angle
of orientation of cylinders with vertical ($\langle \theta \rangle$)
with spherical particle volume fraction ($\phi$) in the region of
interest.  Here, $k_{te0}$ and $k_{re0}$ correspond to the avalanche
for the base case, i.e., in the absence of spherical particles
($\phi = 0.0$). Overall, the total volume fraction (including spheres
as well as cylinders) will vary between the lower limit
[$\langle \phi_{t} \rangle = 0.509$, shown as magenta line in
Fig.~\ref{expt-sim-compr}(a)] in the absence of spherical particles and
upper limit [$\langle \phi_{t} \rangle = 0.59$ shown as orange line in
Fig.~\ref{expt-sim-compr}(a)] in the absence of cylindrical particles (or
only spherical particles). This is not surprising given the fact that
only spheres are expected to pack efficiently than only cylinders.
The average total volume fraction [shown as green squares in
Fig.\ref{expt-sim-compr}(a)] increases quickly for the lower values of
spherical particle concentration followed by a gradual increase at
corresponding higher values.

The initial rapid increase in the average total volume fraction
($\langle \phi_{t} \rangle$) can be envisioned as the spherical
particles filling the available voids between cylinders thereby
improving the packed state. This progressive increase in the value of
volume fraction is expected to gradually decrease the flow
velocity leading to the decrease in the average translational kinetic
energy ($\langle k_{te} \rangle / \langle k_{te0} \rangle$) as indeed
seen in Fig.~\ref{expt-sim-compr}(a). Second, the increased volume
fraction will impede the ability of the cylindrical particles to rotate
and align themselves with the flow direction, thereby reducing the
average rotational kinetic energy
($\langle k_{re} \rangle / \langle k_{re0} \rangle$) as seen in
fig.~\ref{expt-sim-compr}(b). The direct consequence seems to be the
increase in the average orientation angle ($\langle \theta \rangle$)
with respect to the vertical (flow) direction [see
Fig.~\ref{expt-sim-compr}(b)]. The higher the orientation angle of
cylinders with respect to vertical, higher would be the resistance to
flow, while the minimum resistance can be expected when cylinders
align parallel to the flow direction. The combined effect of these
three entities ($k_{te}$, $k_{re}$, and $\theta$) is to make the arch
increasingly resistant to the flow, thereby leading to more frequent
clogging and, hence,  the reduction in avalanche size, in agreement
with the previously published work~\cite{arevalo16}. This is observed
in Fig.~\ref{expt-sim-compr}(b), wherein the decrease in the avalanche
size (shown as green curve) coincides with the decrease in both the
components of kinetic energy [violet curve in
fig.~\ref{expt-sim-compr}(a) and blue curve in
fig.~\ref{expt-sim-compr}(b)] and increase in average orientation angle
with respect to vertical [magenta curve in
Fig.~\ref{expt-sim-compr}(b)].  The minimum avalanche size is obtained
for a spherical particle fraction corresponding to minimum in both
components of kinetic energy, maximum in the orientation angle as well
as the transition between rapid and slow increase in total volume
fraction. This transition point represents the changeover from a state
of higher clogging occurrences to a state of lower clogging
occurrences.

Following the transition point, the total volume fraction increases
gradually with increase in spherical particle fraction. Given that the
total number of voids available are limited and mostly filled up,
further increase in $\phi$ simply adds up the number of spherical
particles leading to a gradual change in total volume fraction. In
view of the argument in the preceding paragraph, this should lead to
further decrease in the value of kinetic energy. On the contrary, both
the components of kinetic energy are seen to increase continuously. As
already discussed, the relative proportion of small sized spherical
particles increases in the material flowing out of the orifice,
leading to a progressive decrease in the mean particle diameter
(number average of cylindrical and spherical particles) with
increasing value of $\phi$. The ratio of the orifice width to this mean
particle diameter increases progressively, thereby reducing the
probability of arch formation, leading to reduced clogging or increased
avalanche size and consequently faster flow, and hence larger
translational kinetic energy. The faster flow, perhaps, enables the
cylinders to rotate more easily and orient themselves with the flow
direction, thereby reducing the angle with the vertical and increase
in the rotational kinetic energy.  The observed non-monotonic
dependence of avalanche size on spherical particle fraction is, then,
due to competing effects between increased packing influencing the
avalanche strength and reduced probability of arch formation with
decreased mean particle size in the outflow zone. Nearly similar
qualitative behavior is also observed for other two size ratios (not
shown).

\section{\label{concl}Conclusions}
The flow of cylindrical particles through a 3-dimensional silo is
investigated in the presence of spherical particles present in
different proportion and of different sizes. Flow visualization
experiments and discrete element method (DEM) simulations are employed
for this study. The clogging behavior is studied for an exit orifice
(or slit) of fixed size and is measured in terms of the size of an
avalanche emanating from the silo.

The presence of spherical particle exhibits a non-monotonicity in the
variation of avalanche size. For small enough spherical particle
fraction, the avalanche size decreases, i.e., the clogging tendency
increases, which is somewhat non-intuitive in nature and in contrast to
previous observations~\cite{nicolas18b,madrid21}. However, for large
enough tracer fraction the avalanche size increases rapidly, i.e., the
clogging tendency decreases in agreement with previous
observations~\cite{nicolas18b,madrid21}. Similar, qualitative,
behavior is observed for all the spherical particle sizes used, though
with certain quantitative differences arising out of size differences.

The non-monotonic behavior of clogging tendency is attributed to two
effects arising due to addition of spherical particles, viz.  increase
in total particle fraction and reduced mean particle size exiting the
orifice. For small enough spherical particle fractions, the former
effect dominates, leading to reduced kinetic energy and
increasingly resistive arch formation. At larger fractions, the latter
effect dominates leading to faster flows, increased kinetic energy,
and reduced clogging tendency.

It is quite interesting to know that such small presence of spheres,
which typically may be neglected, can lead to unexpected clogging,
not quite expected. The knowledge of the existence of such
behavior would be of substantial interest to several industries
handling powders in various applications. Fundamentally, the presence
of such behavior can spur detailed modeling to understand the flow of
bi-disperse granular material through the silo, something which is
rarely studied. More interesting would be to study the carryover of
this phenomena for particles of different shapes and for cohesive
grains encountered in practice.

\begin{acknowledgements}
  A.V.O gratefully acknowledges the financial support from the Science and
  Engineering Research Board, India (Grant No. CRG/2019/000423). S.C.B
  acknowledges the Council of Scientific and Industrial Research
  (CSIR), India, for the CSIR-GATE fellowship. The support and the
  resources provided by ``PARAM Brahma Facility'' under the National
  Supercomputing Mission, Government of India at the Indian Institute
  of Science Education and Research (IISER), Pune is gratefully
  acknowledged. The authors also gratefully acknowledge the
  computational resources provided by ``Einstein cluster facility'' at
  CSIR - National Chemical Laboratory, Pune.
\end{acknowledgements}

\bibliography{ref}

\providecommand{\noopsort}[1]{}\providecommand{\singleletter}[1]{#1}%
\begin{thebibliography}{27}%
\makeatletter
\providecommand \@ifxundefined [1]{%
 \@ifx{#1\undefined}
}%
\providecommand \@ifnum [1]{%
 \ifnum #1\expandafter \@firstoftwo
 \else \expandafter \@secondoftwo
 \fi
}%
\providecommand \@ifx [1]{%
 \ifx #1\expandafter \@firstoftwo
 \else \expandafter \@secondoftwo
 \fi
}%
\providecommand \natexlab [1]{#1}%
\providecommand \enquote  [1]{``#1''}%
\providecommand \bibnamefont  [1]{#1}%
\providecommand \bibfnamefont [1]{#1}%
\providecommand \citenamefont [1]{#1}%
\providecommand \href@noop [0]{\@secondoftwo}%
\providecommand \href [0]{\begingroup \@sanitize@url \@href}%
\providecommand \@href[1]{\@@startlink{#1}\@@href}%
\providecommand \@@href[1]{\endgroup#1\@@endlink}%
\providecommand \@sanitize@url [0]{\catcode `\\12\catcode `\$12\catcode
  `\&12\catcode `\#12\catcode `\^12\catcode `\_12\catcode `\%12\relax}%
\providecommand \@@startlink[1]{}%
\providecommand \@@endlink[0]{}%
\providecommand \url  [0]{\begingroup\@sanitize@url \@url }%
\providecommand \@url [1]{\endgroup\@href {#1}{\urlprefix }}%
\providecommand \urlprefix  [0]{URL }%
\providecommand \Eprint [0]{\href }%
\providecommand \doibase [0]{http://dx.doi.org/}%
\providecommand \selectlanguage [0]{\@gobble}%
\providecommand \bibinfo  [0]{\@secondoftwo}%
\providecommand \bibfield  [0]{\@secondoftwo}%
\providecommand \translation [1]{[#1]}%
\providecommand \BibitemOpen [0]{}%
\providecommand \bibitemStop [0]{}%
\providecommand \bibitemNoStop [0]{.\EOS\space}%
\providecommand \EOS [0]{\spacefactor3000\relax}%
\providecommand \BibitemShut  [1]{\csname bibitem#1\endcsname}%
\let\auto@bib@innerbib\@empty
\bibitem [{\citenamefont {To}, \citenamefont {Lai},\ and\ \citenamefont
  {Pak}(2001)}]{to01}%
  \BibitemOpen
  \bibfield  {author} {\bibinfo {author} {\bibfnamefont {K.}~\bibnamefont
  {To}}, \bibinfo {author} {\bibfnamefont {P.~Y.}\ \bibnamefont {Lai}}, \ and\
  \bibinfo {author} {\bibfnamefont {H.~K.}\ \bibnamefont {Pak}},\ }\href@noop
  {} {\bibfield  {journal} {\bibinfo  {journal} {Phys.\ Rev.\ Lett.}\ }\textbf
  {\bibinfo {volume} {86}},\ \bibinfo {pages} {71} (\bibinfo {year}
  {2001})}\BibitemShut {NoStop}%
\bibitem [{\citenamefont {Tang}\ and\ \citenamefont
  {Behringer}(2011)}]{tang11}%
  \BibitemOpen
  \bibfield  {author} {\bibinfo {author} {\bibfnamefont {J.}~\bibnamefont
  {Tang}}\ and\ \bibinfo {author} {\bibfnamefont {R.~P.}\ \bibnamefont
  {Behringer}},\ }\href@noop {} {\bibfield  {journal} {\bibinfo  {journal}
  {Chaos}\ }\textbf {\bibinfo {volume} {21}},\ \bibinfo {pages} {041107}
  (\bibinfo {year} {2011})}\BibitemShut {NoStop}%
\bibitem [{\citenamefont {Tewari}, \citenamefont {Dichter},\ and\ \citenamefont
  {Chakraborty}(2013)}]{tewari13}%
  \BibitemOpen
  \bibfield  {author} {\bibinfo {author} {\bibfnamefont {S.}~\bibnamefont
  {Tewari}}, \bibinfo {author} {\bibfnamefont {M.}~\bibnamefont {Dichter}}, \
  and\ \bibinfo {author} {\bibfnamefont {B.}~\bibnamefont {Chakraborty}},\
  }\bibfield  {title} {\enquote {\bibinfo {title} {Signatures of incipient
  jamming in collisional hopper flows},}\ }\href@noop {} {\bibfield  {journal}
  {\bibinfo  {journal} {Soft Matter}\ }\textbf {\bibinfo {volume} {9}},\
  \bibinfo {pages} {5016} (\bibinfo {year} {2013})}\BibitemShut {NoStop}%
\bibitem [{\citenamefont {Thomas}\ and\ \citenamefont
  {Durian}(2013)}]{thomas13}%
  \BibitemOpen
  \bibfield  {author} {\bibinfo {author} {\bibfnamefont {C.~C.}\ \bibnamefont
  {Thomas}}\ and\ \bibinfo {author} {\bibfnamefont {D.~J.}\ \bibnamefont
  {Durian}},\ }\bibfield  {title} {\enquote {\bibinfo {title} {Geometry
  dependence of the clogging transition in tilted hoppers},}\ }\href@noop {}
  {\bibfield  {journal} {\bibinfo  {journal} {Phys.\ Rev.\ E}\ }\textbf
  {\bibinfo {volume} {87}},\ \bibinfo {pages} {052201} (\bibinfo {year}
  {2013})}\BibitemShut {NoStop}%
\bibitem [{\citenamefont {Zuriguel}(2014)}]{zuriguel14b}%
  \BibitemOpen
  \bibfield  {author} {\bibinfo {author} {\bibfnamefont {I.}~\bibnamefont
  {Zuriguel}},\ }\bibfield  {title} {\enquote {\bibinfo {title} {Clogging of
  granular material in bottlenecks},}\ }\href@noop {} {\bibfield  {journal}
  {\bibinfo  {journal} {Pap. Phys.}\ }\textbf {\bibinfo {volume} {6}},\
  \bibinfo {pages} {060014} (\bibinfo {year} {2014})}\BibitemShut {NoStop}%
\bibitem [{\citenamefont {Thomas}\ and\ \citenamefont
  {Durian}(2016)}]{thomas16}%
  \BibitemOpen
  \bibfield  {author} {\bibinfo {author} {\bibfnamefont {C.~C.}\ \bibnamefont
  {Thomas}}\ and\ \bibinfo {author} {\bibfnamefont {D.~J.}\ \bibnamefont
  {Durian}},\ }\bibfield  {title} {\enquote {\bibinfo {title} {Intermittency
  and velocity fluctuations in hopper flows prone to clogging},}\ }\href@noop
  {} {\bibfield  {journal} {\bibinfo  {journal} {Phys. Rev. E}\ }\textbf
  {\bibinfo {volume} {94}},\ \bibinfo {pages} {022901} (\bibinfo {year}
  {2016})}\BibitemShut {NoStop}%
\bibitem [{\citenamefont {Janda}\ \emph {et~al.}(2009)\citenamefont {Janda},
  \citenamefont {Maza}, \citenamefont {Garcimart\'{i}n}, \citenamefont {Kolb},
  \citenamefont {Lanuze},\ and\ \citenamefont {Cl\'{e}ment}}]{janda09b}%
  \BibitemOpen
  \bibfield  {author} {\bibinfo {author} {\bibfnamefont {A.}~\bibnamefont
  {Janda}}, \bibinfo {author} {\bibfnamefont {D.}~\bibnamefont {Maza}},
  \bibinfo {author} {\bibfnamefont {A.}~\bibnamefont {Garcimart\'{i}n}},
  \bibinfo {author} {\bibfnamefont {E.}~\bibnamefont {Kolb}}, \bibinfo {author}
  {\bibfnamefont {J.}~\bibnamefont {Lanuze}}, \ and\ \bibinfo {author}
  {\bibfnamefont {E.}~\bibnamefont {Cl\'{e}ment}},\ }\bibfield  {title}
  {\enquote {\bibinfo {title} {Unjamming a granular hopper by vibration},}\
  }\href@noop {} {\bibfield  {journal} {\bibinfo  {journal} {Eur. Phys. Lett.}\
  }\textbf {\bibinfo {volume} {87}},\ \bibinfo {pages} {24002} (\bibinfo {year}
  {2009})}\BibitemShut {NoStop}%
\bibitem [{\citenamefont {Kunte}, \citenamefont {Doshi},\ and\ \citenamefont
  {Orpe}(2014)}]{kunte14}%
  \BibitemOpen
  \bibfield  {author} {\bibinfo {author} {\bibfnamefont {A.}~\bibnamefont
  {Kunte}}, \bibinfo {author} {\bibfnamefont {P.}~\bibnamefont {Doshi}}, \ and\
  \bibinfo {author} {\bibfnamefont {A.~V.}\ \bibnamefont {Orpe}},\ }\bibfield
  {title} {\enquote {\bibinfo {title} {Spontaneous jamming and unjamming in a
  hopper with multiple exit orifices},}\ }\href@noop {} {\bibfield  {journal}
  {\bibinfo  {journal} {Phys.\ Rev.\ E (Rapid Comm.)}\ }\textbf {\bibinfo
  {volume} {90}},\ \bibinfo {pages} {020201(R)} (\bibinfo {year}
  {2014})}\BibitemShut {NoStop}%
\bibitem [{\citenamefont {Orpe}\ and\ \citenamefont {Doshi}(2019)}]{orpe19}%
  \BibitemOpen
  \bibfield  {author} {\bibinfo {author} {\bibfnamefont {A.~V.}\ \bibnamefont
  {Orpe}}\ and\ \bibinfo {author} {\bibfnamefont {P.}~\bibnamefont {Doshi}},\
  }\bibfield  {title} {\enquote {\bibinfo {title} {Friction-mediated flow and
  jamming in a two-dimensional silo with two exit orifices},}\ }\href@noop {}
  {\bibfield  {journal} {\bibinfo  {journal} {Phys. Rev. E}\ }\textbf {\bibinfo
  {volume} {100}},\ \bibinfo {pages} {012901} (\bibinfo {year}
  {2019})}\BibitemShut {NoStop}%
\bibitem [{\citenamefont {Zuriguel}\ \emph {et~al.}(2011)\citenamefont
  {Zuriguel}, \citenamefont {Janda}, \citenamefont {Garcimart\'{i}n},
  \citenamefont {Lozano}, \citenamefont {Ar\'{e}valo},\ and\ \citenamefont
  {Maza}}]{zuriguel11}%
  \BibitemOpen
  \bibfield  {author} {\bibinfo {author} {\bibfnamefont {I.}~\bibnamefont
  {Zuriguel}}, \bibinfo {author} {\bibfnamefont {A.}~\bibnamefont {Janda}},
  \bibinfo {author} {\bibfnamefont {A.}~\bibnamefont {Garcimart\'{i}n}},
  \bibinfo {author} {\bibfnamefont {C.}~\bibnamefont {Lozano}}, \bibinfo
  {author} {\bibfnamefont {R.}~\bibnamefont {Ar\'{e}valo}}, \ and\ \bibinfo
  {author} {\bibfnamefont {D.}~\bibnamefont {Maza}},\ }\bibfield  {title}
  {\enquote {\bibinfo {title} {Silo clogging reduction by the presence of an
  obstacle},}\ }\href@noop {} {\bibfield  {journal} {\bibinfo  {journal}
  {Phys.\ Rev.\ Lett.}\ }\textbf {\bibinfo {volume} {107}},\ \bibinfo {pages}
  {278001} (\bibinfo {year} {2011})}\BibitemShut {NoStop}%
\bibitem [{\citenamefont {Zuriguel}\ \emph {et~al.}(2014)\citenamefont
  {Zuriguel}, \citenamefont {Parisi}, \citenamefont {Hidalgo}, \citenamefont
  {Lozano}, \citenamefont {Janda}, \citenamefont {Gago}, \citenamefont
  {Peralta}, \citenamefont {Ferrer}, \citenamefont {Pugnaloni}, \citenamefont
  {Cl\'{e}ment}, \citenamefont {Maza}, \citenamefont {Pagonabarraga},\ and\
  \citenamefont {Garcimart\'{i}n}}]{zuriguel14a}%
  \BibitemOpen
  \bibfield  {author} {\bibinfo {author} {\bibfnamefont {I.}~\bibnamefont
  {Zuriguel}}, \bibinfo {author} {\bibfnamefont {D.~R.}\ \bibnamefont
  {Parisi}}, \bibinfo {author} {\bibfnamefont {R.~C.}\ \bibnamefont {Hidalgo}},
  \bibinfo {author} {\bibfnamefont {C.}~\bibnamefont {Lozano}}, \bibinfo
  {author} {\bibfnamefont {A.}~\bibnamefont {Janda}}, \bibinfo {author}
  {\bibfnamefont {P.~A.}\ \bibnamefont {Gago}}, \bibinfo {author}
  {\bibfnamefont {J.~P.}\ \bibnamefont {Peralta}}, \bibinfo {author}
  {\bibfnamefont {L.~M.}\ \bibnamefont {Ferrer}}, \bibinfo {author}
  {\bibfnamefont {L.~A.}\ \bibnamefont {Pugnaloni}}, \bibinfo {author}
  {\bibfnamefont {E.}~\bibnamefont {Cl\'{e}ment}}, \bibinfo {author}
  {\bibfnamefont {D.}~\bibnamefont {Maza}}, \bibinfo {author} {\bibfnamefont
  {I.}~\bibnamefont {Pagonabarraga}}, \ and\ \bibinfo {author} {\bibfnamefont
  {A.}~\bibnamefont {Garcimart\'{i}n}},\ }\bibfield  {title} {\enquote
  {\bibinfo {title} {Clogging transition of many-particle systems flowing
  through bottlenecks},}\ }\href@noop {} {\bibfield  {journal} {\bibinfo
  {journal} {Sci. Rep.}\ }\textbf {\bibinfo {volume} {4}},\ \bibinfo {pages}
  {7324} (\bibinfo {year} {2014})}\BibitemShut {NoStop}%
\bibitem [{\citenamefont {Zuriguel}\ \emph {et~al.}(2017)\citenamefont
  {Zuriguel}, \citenamefont {Janda}, \citenamefont {Ar\'{e}valo}, \citenamefont
  {Maza},\ and\ \citenamefont {Garcimart\'{i}n}}]{zuriguel17}%
  \BibitemOpen
  \bibfield  {author} {\bibinfo {author} {\bibfnamefont {I.}~\bibnamefont
  {Zuriguel}}, \bibinfo {author} {\bibfnamefont {A.}~\bibnamefont {Janda}},
  \bibinfo {author} {\bibfnamefont {R.}~\bibnamefont {Ar\'{e}valo}}, \bibinfo
  {author} {\bibfnamefont {D.}~\bibnamefont {Maza}}, \ and\ \bibinfo {author}
  {\bibfnamefont {A.}~\bibnamefont {Garcimart\'{i}n}},\ }\bibfield  {title}
  {\enquote {\bibinfo {title} {Clogging and unclogging of many-particle systems
  passing through a bottleneck},}\ }\href@noop {} {\bibfield  {journal}
  {\bibinfo  {journal} {EPJ Web. Conf.}\ }\textbf {\bibinfo {volume} {140}},\
  \bibinfo {pages} {01002} (\bibinfo {year} {2017})}\BibitemShut {NoStop}%
\bibitem [{\citenamefont {Hidalgo}\ \emph {et~al.}(2018)\citenamefont
  {Hidalgo}, \citenamefont {{n}i Arana}, \citenamefont {Hern\'{a}ndez-Puerta},\
  and\ \citenamefont {Pagonabarraga}}]{hidalgo18}%
  \BibitemOpen
  \bibfield  {author} {\bibinfo {author} {\bibfnamefont {R.~C.}\ \bibnamefont
  {Hidalgo}}, \bibinfo {author} {\bibfnamefont {A.~G.}\ \bibnamefont {{n}i
  Arana}}, \bibinfo {author} {\bibfnamefont {A.}~\bibnamefont
  {Hern\'{a}ndez-Puerta}}, \ and\ \bibinfo {author} {\bibfnamefont
  {I.}~\bibnamefont {Pagonabarraga}},\ }\bibfield  {title} {\enquote {\bibinfo
  {title} {Flow of colloidal suspensions through small orifices},}\ }\href@noop
  {} {\bibfield  {journal} {\bibinfo  {journal} {Phys. Rev. E}\ }\textbf
  {\bibinfo {volume} {97}},\ \bibinfo {pages} {012611} (\bibinfo {year}
  {2018})}\BibitemShut {NoStop}%
\bibitem [{\citenamefont {Nicolas}, \citenamefont {Kuperman},\ and\
  \citenamefont {Bouzat}(2018)}]{nicolas18b}%
  \BibitemOpen
  \bibfield  {author} {\bibinfo {author} {\bibfnamefont {A.}~\bibnamefont
  {Nicolas}}, \bibinfo {author} {\bibfnamefont {M.~N.}\ \bibnamefont
  {Kuperman}}, \ and\ \bibinfo {author} {\bibfnamefont {S.}~\bibnamefont
  {Bouzat}},\ }\bibfield  {title} {\enquote {\bibinfo {title} {A
  counterintuitive way to speed up pedestrian and granular bottleneck flows
  prone to clogging: can 'more' escape faster},}\ }\href@noop {} {\bibfield
  {journal} {\bibinfo  {journal} {J. Stat. Mech.}\ ,\ \bibinfo {pages}
  {083403}} (\bibinfo {year} {2018})}\BibitemShut {NoStop}%
\bibitem [{\citenamefont {Madrid}\ \emph {et~al.}(2021)\citenamefont {Madrid},
  \citenamefont {Carlevaro}, \citenamefont {Pugnaloni}, \citenamefont
  {Kuperman},\ and\ \citenamefont {Bouzat}}]{madrid21}%
  \BibitemOpen
  \bibfield  {author} {\bibinfo {author} {\bibfnamefont {M.~A.}\ \bibnamefont
  {Madrid}}, \bibinfo {author} {\bibfnamefont {C.~M.}\ \bibnamefont
  {Carlevaro}}, \bibinfo {author} {\bibfnamefont {L.~A.}\ \bibnamefont
  {Pugnaloni}}, \bibinfo {author} {\bibfnamefont {M.}~\bibnamefont {Kuperman}},
  \ and\ \bibinfo {author} {\bibfnamefont {S.}~\bibnamefont {Bouzat}},\
  }\bibfield  {title} {\enquote {\bibinfo {title} {Enhancement of the flow of
  vibrated grains through narrow apertures by addition of small particles},}\
  }\href@noop {} {\bibfield  {journal} {\bibinfo  {journal} {Phys. Rev. E}\
  }\textbf {\bibinfo {volume} {103}},\ \bibinfo {pages} {L030901} (\bibinfo
  {year} {2021})}\BibitemShut {NoStop}%
\bibitem [{\citenamefont {Zuriguel}\ \emph {et~al.}(2005)\citenamefont
  {Zuriguel}, \citenamefont {Garcimart\'{i}n}, \citenamefont {Maza},
  \citenamefont {Pugnaloni},\ and\ \citenamefont {Pastor}}]{zuriguel05}%
  \BibitemOpen
  \bibfield  {author} {\bibinfo {author} {\bibfnamefont {I.}~\bibnamefont
  {Zuriguel}}, \bibinfo {author} {\bibfnamefont {A.}~\bibnamefont
  {Garcimart\'{i}n}}, \bibinfo {author} {\bibfnamefont {D.}~\bibnamefont
  {Maza}}, \bibinfo {author} {\bibfnamefont {L.~A.}\ \bibnamefont {Pugnaloni}},
  \ and\ \bibinfo {author} {\bibfnamefont {J.~M.}\ \bibnamefont {Pastor}},\
  }\bibfield  {title} {\enquote {\bibinfo {title} {Jamming during the discharge
  of granular matter from a silo},}\ }\href@noop {} {\bibfield  {journal}
  {\bibinfo  {journal} {Phys. Rev. E}\ }\textbf {\bibinfo {volume} {71}},\
  \bibinfo {pages} {051303} (\bibinfo {year} {2005})}\BibitemShut {NoStop}%
\bibitem [{\citenamefont {Goldberg}, \citenamefont {Carlevaro},\ and\
  \citenamefont {Pugnaloni}(2015)}]{goldberg15}%
  \BibitemOpen
  \bibfield  {author} {\bibinfo {author} {\bibfnamefont {E.}~\bibnamefont
  {Goldberg}}, \bibinfo {author} {\bibfnamefont {C.~M.}\ \bibnamefont
  {Carlevaro}}, \ and\ \bibinfo {author} {\bibfnamefont {L.~A.}\ \bibnamefont
  {Pugnaloni}},\ }\bibfield  {title} {\enquote {\bibinfo {title} {Flow rate of
  polygonal grains through a bottleneck: Interplay between shape and size},}\
  }\href@noop {} {\bibfield  {journal} {\bibinfo  {journal} {Pap. Phys.}\
  }\textbf {\bibinfo {volume} {7}},\ \bibinfo {pages} {070016} (\bibinfo {year}
  {2015})}\BibitemShut {NoStop}%
\bibitem [{\citenamefont {Goldberg}, \citenamefont {Carlevaro},\ and\
  \citenamefont {Pugnaloni}(2017)}]{goldberg17}%
  \BibitemOpen
  \bibfield  {author} {\bibinfo {author} {\bibfnamefont {E.}~\bibnamefont
  {Goldberg}}, \bibinfo {author} {\bibfnamefont {C.~M.}\ \bibnamefont
  {Carlevaro}}, \ and\ \bibinfo {author} {\bibfnamefont {L.~A.}\ \bibnamefont
  {Pugnaloni}},\ }\bibfield  {title} {\enquote {\bibinfo {title} {Effect of
  grain shape on the jamming of two-dimensional silos},}\ }\href@noop {}
  {\bibfield  {journal} {\bibinfo  {journal} {EPJ Web. Conf.}\ }\textbf
  {\bibinfo {volume} {140}},\ \bibinfo {pages} {06009} (\bibinfo {year}
  {2017})}\BibitemShut {NoStop}%
\bibitem [{\citenamefont {Thornto}, \citenamefont {Cummins},\ and\
  \citenamefont {Cleary}(2011)}]{thornton11}%
  \BibitemOpen
  \bibfield  {author} {\bibinfo {author} {\bibfnamefont {C.}~\bibnamefont
  {Thornto}}, \bibinfo {author} {\bibfnamefont {S.~J.}\ \bibnamefont
  {Cummins}}, \ and\ \bibinfo {author} {\bibfnamefont {P.~W.}\ \bibnamefont
  {Cleary}},\ }\bibfield  {title} {\enquote {\bibinfo {title} {An investigation
  of the comparative behaviour of alternative contact force models during
  elastic collisions},}\ }\href@noop {} {\bibfield  {journal} {\bibinfo
  {journal} {Powder Technol.}\ }\textbf {\bibinfo {volume} {210}},\ \bibinfo
  {pages} {189--197} (\bibinfo {year} {2011})}\BibitemShut {NoStop}%
\bibitem [{\citenamefont {Yeom}\ \emph {et~al.}(2019)\citenamefont {Yeom},
  \citenamefont {Ha}, \citenamefont {Kim}, \citenamefont {an~S.~Hwang},\ and\
  \citenamefont {Choi}}]{yeom19}%
  \BibitemOpen
  \bibfield  {author} {\bibinfo {author} {\bibfnamefont {S.~B.}\ \bibnamefont
  {Yeom}}, \bibinfo {author} {\bibfnamefont {E.}~\bibnamefont {Ha}}, \bibinfo
  {author} {\bibfnamefont {M.}~\bibnamefont {Kim}}, \bibinfo {author}
  {\bibfnamefont {S.~H.~J.}\ \bibnamefont {an~S.~Hwang}}, \ and\ \bibinfo
  {author} {\bibfnamefont {D.~H.}\ \bibnamefont {Choi}},\ }\bibfield  {title}
  {\enquote {\bibinfo {title} {Application of the discrete element method for
  manufacturing process simulation in the pharmaceutical industry},}\
  }\href@noop {} {\bibfield  {journal} {\bibinfo  {journal} {Pharmaceutics}\
  }\textbf {\bibinfo {volume} {11}},\ \bibinfo {pages} {414} (\bibinfo {year}
  {2019})}\BibitemShut {NoStop}%
\bibitem [{\citenamefont {Markauskas}\ and\ \citenamefont
  {Kacianauskas}(2011)}]{markaus11}%
  \BibitemOpen
  \bibfield  {author} {\bibinfo {author} {\bibfnamefont {D.}~\bibnamefont
  {Markauskas}}\ and\ \bibinfo {author} {\bibfnamefont {R.}~\bibnamefont
  {Kacianauskas}},\ }\bibfield  {title} {\enquote {\bibinfo {title}
  {Investigation of rice grain flow by multi-sphere particle model with rolling
  resistance},}\ }\href@noop {} {\bibfield  {journal} {\bibinfo  {journal}
  {Gran. Mat.}\ }\textbf {\bibinfo {volume} {13}},\ \bibinfo {pages} {143--148}
  (\bibinfo {year} {2011})}\BibitemShut {NoStop}%
\bibitem [{\citenamefont {Liu}\ \emph {et~al.}(2014)\citenamefont {Liu},
  \citenamefont {Zhou}, \citenamefont {Zou}, \citenamefont {Pinson},\ and\
  \citenamefont {Yu}}]{liu14}%
  \BibitemOpen
  \bibfield  {author} {\bibinfo {author} {\bibfnamefont {S.~D.}\ \bibnamefont
  {Liu}}, \bibinfo {author} {\bibfnamefont {Z.~Y.}\ \bibnamefont {Zhou}},
  \bibinfo {author} {\bibfnamefont {R.~P.}\ \bibnamefont {Zou}}, \bibinfo
  {author} {\bibfnamefont {D.}~\bibnamefont {Pinson}}, \ and\ \bibinfo {author}
  {\bibfnamefont {A.~B.}\ \bibnamefont {Yu}},\ }\bibfield  {title} {\enquote
  {\bibinfo {title} {Flow characteristics and discharge rate of ellipsoidal
  particles in a flat bottom hopper},}\ }\href@noop {} {\bibfield  {journal}
  {\bibinfo  {journal} {Powder Technol.}\ }\textbf {\bibinfo {volume} {253}},\
  \bibinfo {pages} {70--79} (\bibinfo {year} {2014})}\BibitemShut {NoStop}%
\bibitem [{\citenamefont {Tangri}, \citenamefont {Guo},\ and\ \citenamefont
  {Curtis}(2019)}]{tangri19}%
  \BibitemOpen
  \bibfield  {author} {\bibinfo {author} {\bibfnamefont {H.}~\bibnamefont
  {Tangri}}, \bibinfo {author} {\bibfnamefont {Y.}~\bibnamefont {Guo}}, \ and\
  \bibinfo {author} {\bibfnamefont {J.~S.}\ \bibnamefont {Curtis}},\ }\bibfield
   {title} {\enquote {\bibinfo {title} {Hopper discharge of elongated particles
  of varying aspect ratio: Experiments and dem simulations},}\ }\href@noop {}
  {\bibfield  {journal} {\bibinfo  {journal} {Chem. Eng. Sci.}\ }\textbf
  {\bibinfo {volume} {4}},\ \bibinfo {pages} {100040} (\bibinfo {year}
  {2019})}\BibitemShut {NoStop}%
\bibitem [{\citenamefont {Rycroft}, \citenamefont {Orpe},\ and\ \citenamefont
  {Kudrolli}(2009)}]{rycroft09}%
  \BibitemOpen
  \bibfield  {author} {\bibinfo {author} {\bibfnamefont {C.~H.}\ \bibnamefont
  {Rycroft}}, \bibinfo {author} {\bibfnamefont {A.~V.}\ \bibnamefont {Orpe}}, \
  and\ \bibinfo {author} {\bibfnamefont {A.}~\bibnamefont {Kudrolli}},\
  }\href@noop {} {\bibfield  {journal} {\bibinfo  {journal} {Phys.\ Rev.\ E}\
  }\textbf {\bibinfo {volume} {80}},\ \bibinfo {pages} {031305} (\bibinfo
  {year} {2009})}\BibitemShut {NoStop}%
\bibitem [{\citenamefont {Ar\'{e}valo}\ and\ \citenamefont
  {Zuriguel}(2016)}]{arevalo16}%
  \BibitemOpen
  \bibfield  {author} {\bibinfo {author} {\bibfnamefont {R.}~\bibnamefont
  {Ar\'{e}valo}}\ and\ \bibinfo {author} {\bibfnamefont {I.}~\bibnamefont
  {Zuriguel}},\ }\bibfield  {title} {\enquote {\bibinfo {title} {Clogging of
  granular materials in silos: effect of gravity and outlet size},}\
  }\href@noop {} {\bibfield  {journal} {\bibinfo  {journal} {Soft Matter}\
  }\textbf {\bibinfo {volume} {12}},\ \bibinfo {pages} {123} (\bibinfo {year}
  {2016})}\BibitemShut {NoStop}%
\bibitem [{\citenamefont {Silbert}\ \emph {et~al.}(2001)\citenamefont
  {Silbert}, \citenamefont {Ertas}, \citenamefont {Grest}, \citenamefont
  {Halsey}, \citenamefont {Levine},\ and\ \citenamefont
  {Plimpton}}]{silbert01}%
  \BibitemOpen
  \bibfield  {author} {\bibinfo {author} {\bibfnamefont {L.~E.}\ \bibnamefont
  {Silbert}}, \bibinfo {author} {\bibfnamefont {D.}~\bibnamefont {Ertas}},
  \bibinfo {author} {\bibfnamefont {G.~S.}\ \bibnamefont {Grest}}, \bibinfo
  {author} {\bibfnamefont {T.~C.}\ \bibnamefont {Halsey}}, \bibinfo {author}
  {\bibfnamefont {D.}~\bibnamefont {Levine}}, \ and\ \bibinfo {author}
  {\bibfnamefont {S.~J.}\ \bibnamefont {Plimpton}},\ }\bibfield  {title}
  {\enquote {\bibinfo {title} {Granular flow down an inclined plane: Bagnold
  scaling and rheology},}\ }\href@noop {} {\bibfield  {journal} {\bibinfo
  {journal} {Phys. Rev. E}\ }\textbf {\bibinfo {volume} {64}},\ \bibinfo
  {pages} {051302} (\bibinfo {year} {2001})}\BibitemShut {NoStop}%
\bibitem [{\citenamefont {Landry}\ \emph {et~al.}(2003)\citenamefont {Landry},
  \citenamefont {Grest}, \citenamefont {Silbert},\ and\ \citenamefont
  {Plimpton}}]{landry03}%
  \BibitemOpen
  \bibfield  {author} {\bibinfo {author} {\bibfnamefont {J.~W.}\ \bibnamefont
  {Landry}}, \bibinfo {author} {\bibfnamefont {G.~S.}\ \bibnamefont {Grest}},
  \bibinfo {author} {\bibfnamefont {L.~E.}\ \bibnamefont {Silbert}}, \ and\
  \bibinfo {author} {\bibfnamefont {S.~J.}\ \bibnamefont {Plimpton}},\
  }\href@noop {} {\bibfield  {journal} {\bibinfo  {journal} {Phys.\ Rev.\ E}\
  }\textbf {\bibinfo {volume} {67}},\ \bibinfo {pages} {041303} (\bibinfo
  {year} {2003})}\BibitemShut {NoStop}%
\end{thebibliography}%

\end{document}